\documentclass[reprint,amsmath,amssymb,aps,pra,onecolumn]{revtex4-2}

\usepackage{graphicx}
\usepackage{bm}

\usepackage{makecell}

\begin{document}

\preprint{APS/123-QED}

\title{Bounded confidence model on growing populations}

\author{Yérali Gandica}
 \homepage{ygandica@gmail.com}
\affiliation{
 Department of Mathematics. Valencian International University, Spain.}%

\author{Guillaume Deffuant}
\affiliation{%
 Université Clermont-Auvergne, INRAE, UR LISC, France.}%

\date{\today}

\begin{abstract}
This paper studies the bounded confidence model on growing fully-mixed populations. In this model, in addition to the usual opinion clusters, significant secondary clusters of smaller size appear systematically, while those secondary clusters appear erratically and include much fewer agents when the population is fixed. Through simulations, we derive the bifurcation diagram of the growing population model and compare it to the diagram obtained with an evolving probability density instead of agents, and with their equivalent with a fixed population. Our tests when changing the usual bounded confidence function into a smooth  bounded confidence function suggest that these secondary clusters are mainly generated by a different mechanism when the population is growing than when it is fixed. 

\keywords{Opinion dynamics \and Bounded confidence \and Minor clusters \and Growing population \and Density model}
\end{abstract}

\maketitle

\section{Introduction}
\label{intro}
Opinion dynamics models express mathematically some hypotheses about social interactions and provide means to investigate their effect on large populations of interacting agents. In particular, bounded confidence models  \cite{Deffuant2000,Hegselmann2002}, assume that when an agent's opinion is too far from the one of its interlocutor, it has no influence. This assumption is grounded in well established results in social psychology \cite{Hovland1980}. Running this model on a population where all the agents hold the same confidence bound, leads to one or more separate clusters of opinions, depending on the value of the confidence bound. Many papers are devoted to studying these models and their variants \cite{Lorenz2007}. Early papers study the model in which all agents have the same confidence bound, showing the details of the bifurcation diagrams when model parameters change. Other researchers investigate populations with different confidence bounds. Several studies focus on including so-called extremists agents, whose opinion is at the border of the opinion interval \cite{Deffuant2002,Deffuant2006b,Mathias2016}. Others consider agents with confidences drawn in a given interval\cite{Schawe2021}. Many papers assess the effect of different types of networks of interactions \cite{Lorenz2007a,Stauffer2004,Gargulio2017}, and in some of them the topology is allowed to change in terms of the opinion dynamics \cite{Carletti2011,Kan2022}. Introducing noise in these models also significantly modifies their qualitative behaviour \cite{Pineda2009,Pineda2011,Flache2011}. Several review papers are totally or partly devoted to these models \cite{Lorenz2007,Castellano2007,Flache2017}.

In this paper, we study the bounded confidence model when new agents are progressively added to the population. As far as we know, the model has not been studied in these conditions yet. The growing population can be related to online communities of agents that appear and then may grow more or less rapidly. Recent models inspired by the physics of gels address, more specifically, the dynamics of aggregation and desegregation of online groups \cite{Manrique2018}. In this paper, we focus on the simple case of a fully mixed population. 

The main new feature appearing when the population is growing is the emergence of significant and stable secondary clusters, located approximately at one confidence bound distance of the main clusters. Secondary or minor clusters were already identified in the fixed population model, more particularly in the density version of the model \cite{BenNaim2003}. However, in simulations of the agent models, the secondary clusters do not appear systematically and when they do, they generally include very few isolated agents. Hence the secondary clusters are often ignored in papers about agent models. However, when the population is growing, it is impossible to ignore them as they appear systematically and include significant numbers of agents, though remaining much smaller than the number of agents in the primary clusters.

These observations led us to question the origin of the secondary clusters. With this aim, we also run simulations of the model with a smooth influence function. Actually, as noticed by several scholars \cite{Deffuant2002,Deffuant2006b}, the bounded confidence influence function shows a strong discontinuity when the distance of the opinions are around the confidence bound, which seems difficult to justify psychologically. We propose a smooth influence function that eliminates this discontinuity and is simpler than the previous versions of smooth bounded confidence influence functions \cite{Deffuant2002,Deffuant2006b}. The analysis using the smooth influence function helps us to uncover that the main mechanism responsible for the systematic presence of secondary clusters in the growing version is the arrival of new agents in regions that are already free of attraction by the primary clusters while the secondary clusters appearing in the model with a fixed population are due to the discontinuity in the interaction function. 

The next section presents the model in all the versions considered in the manuscript: agent-based, density-based, growing or fixed population, and finally, the smooth influence function. The following section is devoted to the simulation results, where we compare the models in detail, showing their similarities and differences. The final section proposes a discussion of these results.

\section{The agent model and its continuous version}

\subsection{Agent model with standard influence}

We consider a population of number $N(t)$ of agents growing with time $t$, with $N(0) = N_0$. An agent $i \in \{1,..,N(t)\}$ is characterised by an opinion $a_i(t) \in [0,1]$. All the agents share the same confidence bound $\epsilon$. At each time step, we perform the classical bounded confidence interaction as in \cite{Deffuant2000}. Moreover, after each interaction, with a probability $\frac{\omega N_0 }{N(t)}$, $N(t)$ being the current population size and $\omega \in [0,1]$ being a parameter, a new agent is added to the population with an opinion uniformly drawn in the opinion interval $[0,1]$. More precisely, the algorithm is as follows.

\begin{enumerate}
    \item Set $N_m$ the maximum number of agents in the model;
    \item Set $t = 0$;
    \item Initialise $N_0$ agents, with opinions uniformly drawn in $[0, 1]$;
    \item While $N_t < N_m$ do:

        \begin{enumerate}
        \item Choose two distinct agents $i$ and $j$ at random and:
        \begin{align}
    \mbox{If } |a_i(t) - a_j(t) | < \epsilon \mbox{ then }
    \begin{cases}
    a_i(t+1) = a_i(t) + \mu(a_j(t) - a_i(t)),\\
    a_j(t+1) = a_j(t) + \mu(a_i(t) - a_j(t)),
    \end{cases}
\end{align}

where $\mu$ is a parameter of the model, fixed to $0.5$ in our simulations;
    \item With probability $\frac{\omega N_0}{N_t}$, add a new agent to the population with an opinion uniformly drawn in $[0, 1]$;
    \item $t := t+1$,  $N_t :=$ size of population.
    \end{enumerate}

\end{enumerate}

As a result, during $N_t$ interactions, where each agent of the population interacts once on average, the population grows of $\omega N_0$ agents on average. The rationale is that the flow of incoming agents is constant during the time needed for each agent to interact once on average. In the following, we refer to iterations as periods of $N_t$ interactions.

\subsection{Density model}

Developing a density model approximating the agent version is a classical practice in agent-based modelling (see, for instance, \cite{BenNaim2003,Deffuant2006b}). The principle is to consider the evolution of the distribution for the probability of the presence of agents instead of considering the agents themselves. The density model can be seen as an agent model with an infinite number of agents starting from a perfectly uniform distribution and synchronous interactions. Comparing agent and density models allows identifying the effects of irregularities and noise with respect to an ideal case.

In practice, to run the evolution of the density distribution numerically, it is necessary to cut the opinion axis into a large number $M$ of intervals and consider the density of agents in each of these intervals. Therefore, the system's state is a vector $d(t) = (d_1,..., d_M)$ of $M$ continuous values, approximating the continuous density. The algorithm is based on the agent model rules and computes the probabilities that the density increases or decreases in each interval. This is generally done through a master equation expressing the inflow and outflow sum at each interval. The repeated action of these changes at each interval produces the evolution of the density.

Here we use an algorithm for applying the master equation by averaging the dynamics of the model and storing all the time-step changes into a vector denoted by $\delta$. Then, all the changes of the density for an iteration are performed at once by adding $\delta$ to the current density $d(t)$. Then, we perform the addition of $\omega N_0$ agents to a population. As the sum of $d_i(t) + \delta_i$, for all the intervals $i \in (1,...,M)$, is 1, we add the vector $(\frac{\omega}{Mt},...,\frac{\omega}{Mt})$. Then we normalise $d(t+1)$. Finally, after the $T$ iterations, we multiply $d(t)$  by $\frac{1 + \omega t}{1 + \omega T}$ for $t \in (0,...,T)$, to get growing values of $d(t)$ like in the agent model.

The algorithm is the following:

\begin{enumerate}
    \item Initialise $d(0) = (\frac{1}{M},...,\frac{1}{M})$, $t = 0$;
    \item For $t \in (0,...,T-1)$:
    \begin{enumerate}
        \item Initialise $\delta = (0,...,0)$ vector of size $M$;
        \item For $i \in (1,...,M)$
        \begin{itemize}
            \item For $j \in (1,...,M)$
            \begin{itemize}
                \item if $|i - j| < \epsilon M$,
                \begin{itemize}
                    \item $k = \mbox{round}(i + \mu(j - i))$;
                    \item $\delta_i := \delta_i - d_i(t) d_j(t)$;
                    \item $\delta_k := \delta_k + d_i(t) d_j(t)$;
                \end{itemize}
            \end{itemize}
        \end{itemize}
        \item For $i \in (1,...,M)$, $d_i(t+1) = d_i(t) + \delta_i + \frac{\omega}{M(1 + \omega(t-1)) }$;
        \item $d(t+1) := \frac{d(t+1)}{\sum_{i=1}^M d_i(t+1)}$
    \end{enumerate}
\item  For $t \in (0,...,T)$, $d(t) = \frac{1 + \omega t}{1 + \omega T} d(t) $
\end{enumerate}

\subsection{Agent model with smooth influence}

With the standard influence rule, the influence of agent $i$ on agent $j$ is strongly discontinuous when the distance of the opinions $|d_{ij}(t)| = |a_i(t) - a_j(t)|$ is around $\epsilon$. Indeed, if $|d_{ij}(t)|$ is slightly below $\epsilon$, the change of opinion $|a_j(t+1) - a_j(t)|$ is close to $\mu \epsilon$, which is the maximum possible change. However, if $|d_{ij}(t)|$ reaches $\epsilon$, the change suddenly drops to 0.

This discontinuity seems difficult to justify psychologically to the eyes of some authors who proposed some changes in the influence function that eliminate this discontinuity (see e.g. \cite{Deffuant2002,Deffuant2006b}). In this paper, we propose the following variant, with $d_{ij}(t) = a_i(t) - a_j(t)$:

\begin{align}
    \mbox{If } |d_{ij}(t) | < \epsilon \mbox{ then }
    \begin{cases}
    a_i(t+1) = a_i(t) + \mu d_{ji}(t)(\epsilon - |d_{ij}(t)|),\\
    a_j(t+1) = a_j(t) + \mu d_{ij}(t)(\epsilon - |d_{ij}(t)|),
    \end{cases}
\end{align}

We refer to this version as the smooth influence and to the previous version as the standard influence. The smooth influence can, of course, be introduced in both versions: the agent model and its continuous version. Here we will limit ourselves to the first case to make a slight comparison.

\section{Simulations with the standard influence function}
\subsection{Examples for $N_0 = 1000$ and $N_0 = 5$}

\begin{figure}[tp]
    \centering
   Confidence bound $\epsilon = 0.3$, initial agent number $N_0 = 1000$
    \begin{tabular}{m{0.3cm}ccp{0.3cm}}
    & & & \\
    & Growing ($\omega = 0.01$)   &  Not growing ($\omega = 0)$ &  \\
     \rotatebox{90}{\hspace{0.0 cm} Opinions} & \makecell{\includegraphics[width = 6 cm]{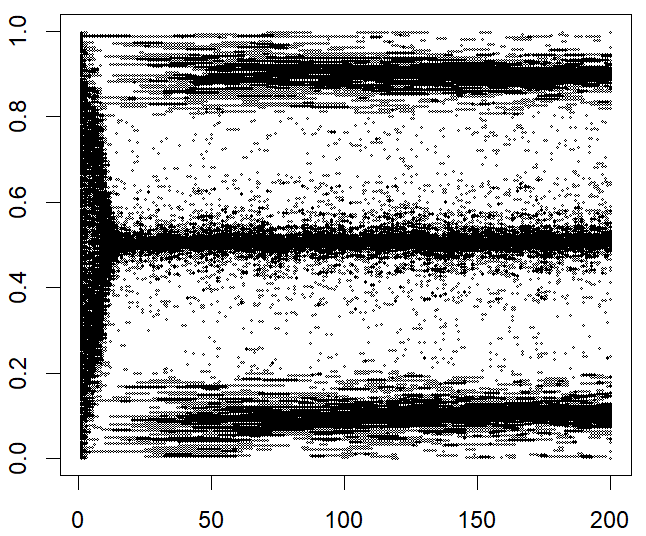}}    &  \makecell{\includegraphics[width = 6 cm]{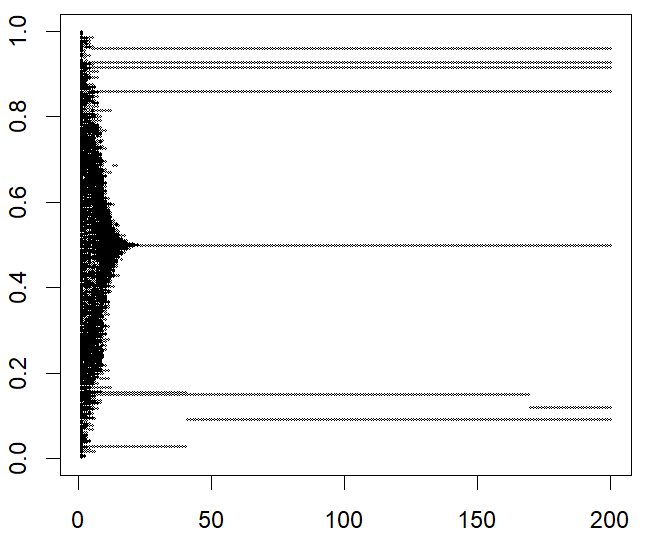}} &  \rotatebox{90}{\hspace{0.0 cm} Agents  }\\

      \rotatebox{90}{\hspace{0.0 cm}  Opinions} & \makecell{\includegraphics[width = 6 cm]{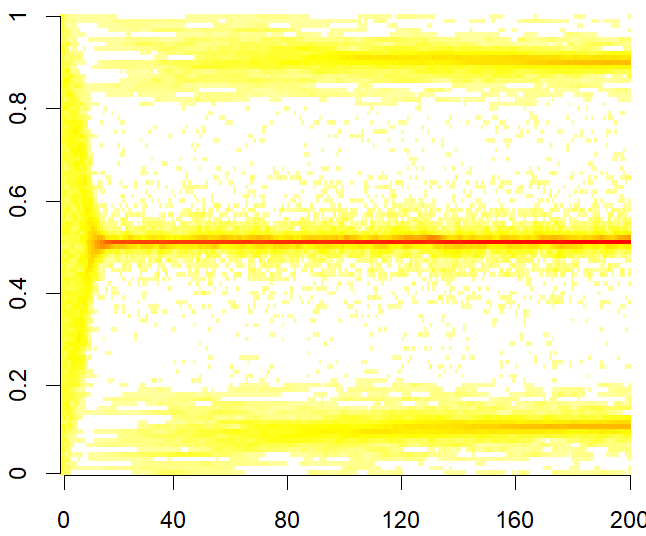}}    &   \makecell{\includegraphics[width = 6 cm]{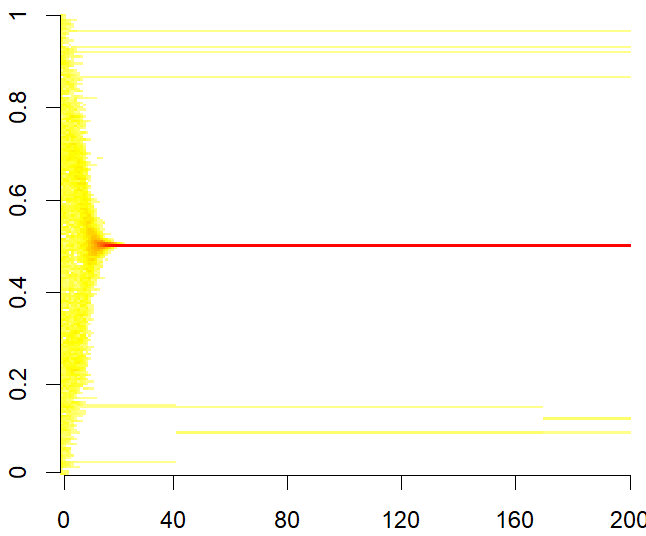}} & \rotatebox{90}{Opinion density } \\

       \rotatebox{90}{\hspace{0.0 cm}  Opinions} & \makecell{\includegraphics[width = 6 cm]{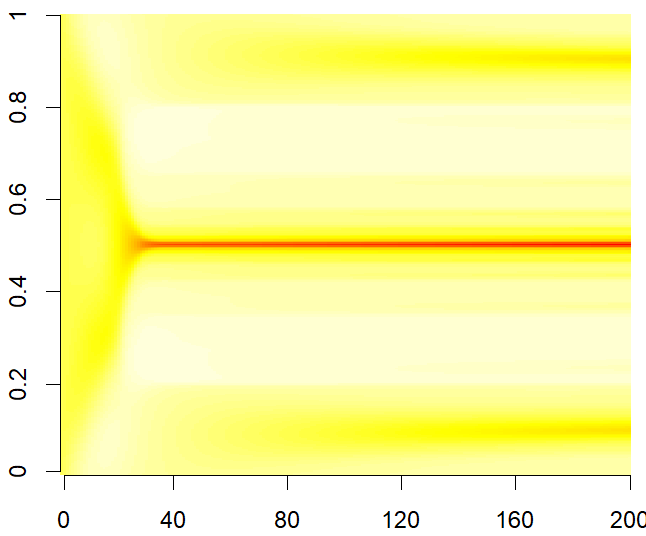}}    &   \makecell{\includegraphics[width = 6 cm]{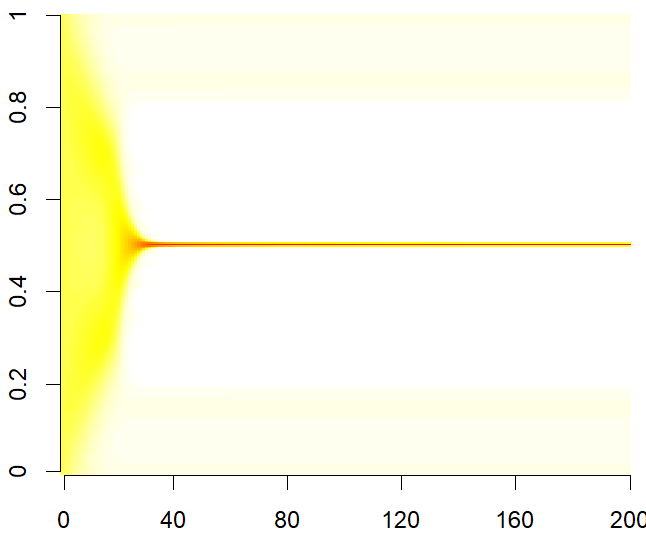}} & \rotatebox{90}{Density model} \\
     &  $t$ / $N_t$ & $t$ / $N_t$ &\\
    \end{tabular}

    \caption{Comparing growing and non-growing cases for the confidence bound $\epsilon = 0.3$. First line of panels represents directly the opinions, the second and third lines represent densities of opinions. The second line represents the density of opinions in the agent model and the third line is directly the output of the density model (see text for details). }
    \label{fig:examples03}
\end{figure}

\begin{figure}[tp]
    \centering
   Confidence bound $\epsilon = 0.2$, initial agent number $N_0 = 1000$

    \begin{tabular}{m{0.3cm}ccm{0.3cm}}
    & & & \\
    & Growing ($\omega = 0.01$)   &  Not growing ($\omega = 0)$ &  \\
     \rotatebox{90}{\hspace{0 cm} Opinions} & \makecell{\includegraphics[width = 6 cm]{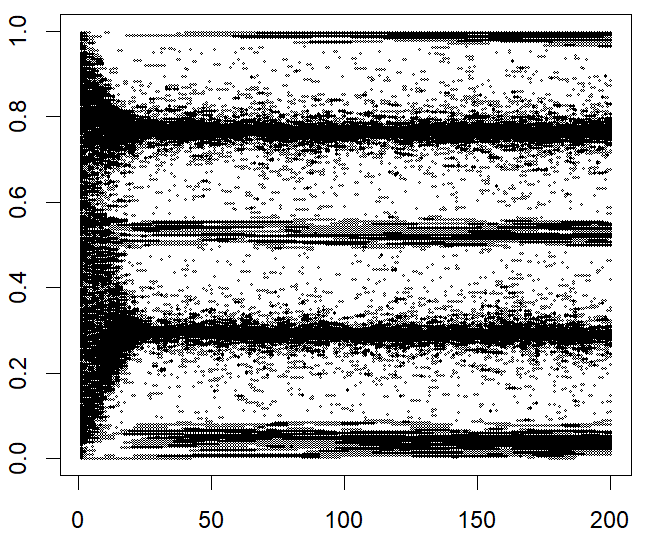}}    &  \makecell{\includegraphics[width = 6 cm]{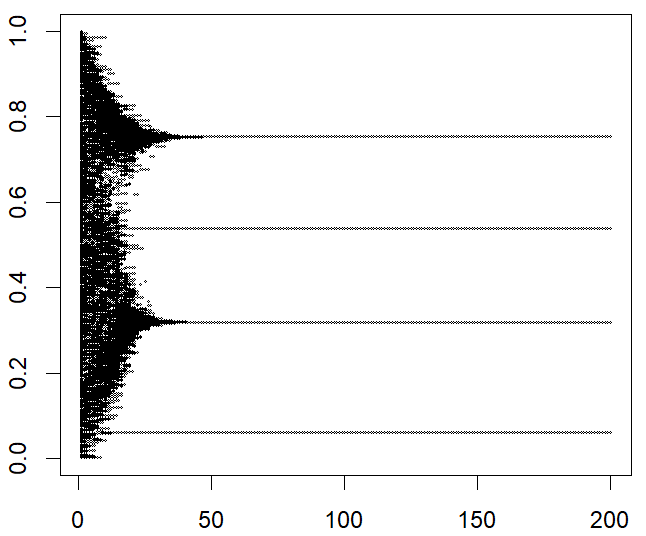}} &  \rotatebox{90}{\hspace{0 cm} Agents }\\

      \rotatebox{90}{\hspace{0 cm}  Opinions} & \makecell{\includegraphics[width = 6 cm]{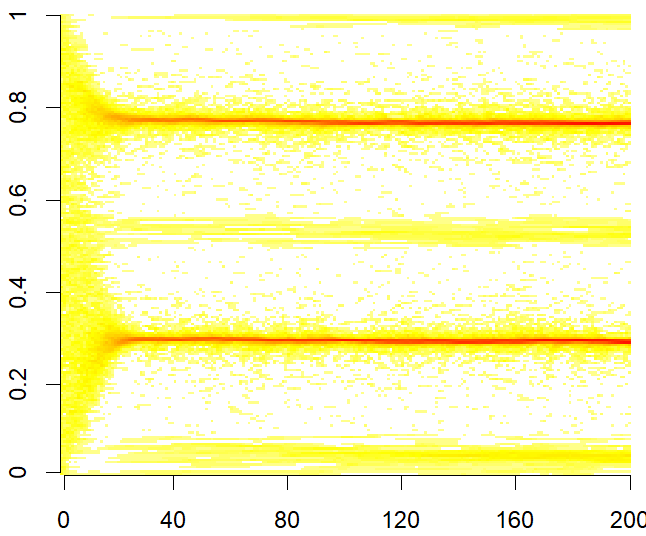}}    &   \makecell{\includegraphics[width = 6 cm]{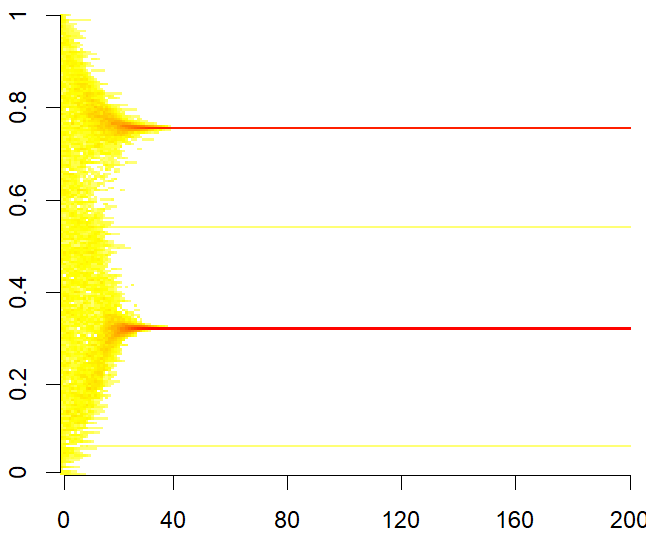}} & \rotatebox{90}{\hspace{0 cm} Opinion density} \\

       \rotatebox{90}{\hspace{0 cm}  Opinions} & \makecell{\includegraphics[width = 6 cm]{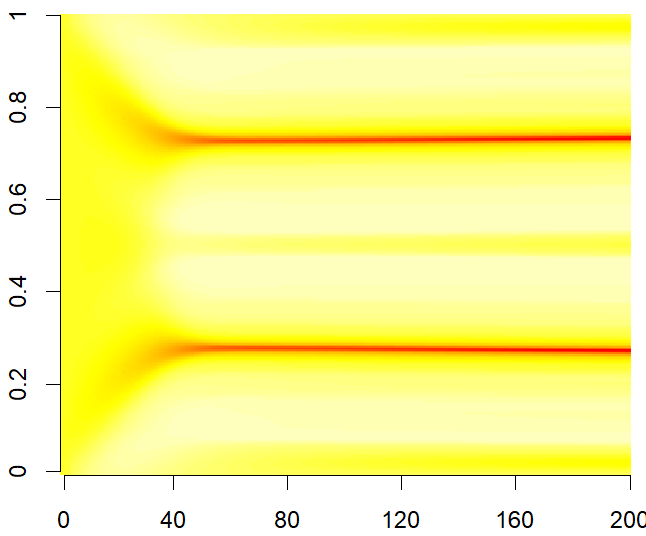}}    &   \makecell{\includegraphics[width = 6 cm]{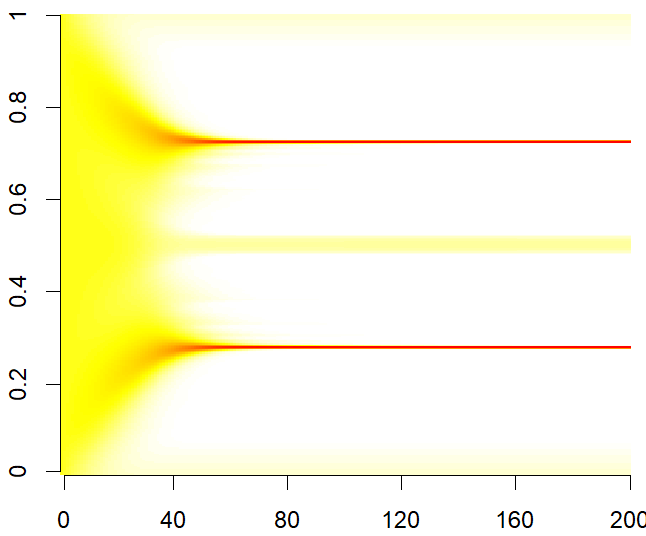}} & \rotatebox{90}{\hspace{0 cm} Density model} \\
     &  $t$ / $N_t$ & $t$ / $N_t$ &\\
    \end{tabular}

    \caption{Comparing growing and non-growing cases for the confidence bound $\epsilon = 0.2$. The first line of panels represents directly the opinions, the second and third lines represent densities of opinions. The second line represents the density of opinions in the agent model and the third line is directly the output of the density model (see text for details). }
    \label{fig:examples02}
\end{figure}

Figures \ref{fig:examples03} and \ref{fig:examples02} compare the model simulations with the standard influence function, for both the agent and the density models. The second rows of panels represents the density of opinions instead of the opinions directly, for the same run as the one represented on the first line of panels. The density of opinions is obtained by counting the number of opinions in 200 regular intervals of the opinion axis. This number is divided by the total number of agents at the end of the simulation.

The main observations from these first examples are the following:
\begin{itemize}
    \item In the growing population model, the secondary clusters are significantly larger than in the non-growing case.
    \item In the growing model, there is a density of agents remaining around the major clusters, while this density is null around the major clusters in the non-growing model.
    \item In general, the results of the agent and density models are close.
\end{itemize}

\begin{figure}[tp]
    \centering
   Initial agent number $N_0 = 5$, growing $\omega = 1$

    \begin{tabular}{m{0.3cm}ccm{0.3cm}}
    & & & \\
    &  $\epsilon = 0.3$   &  $\epsilon = 0.2$ &  \\
     \rotatebox{90}{\hspace{0 cm} Opinions} & \makecell{\includegraphics[width = 5 cm]{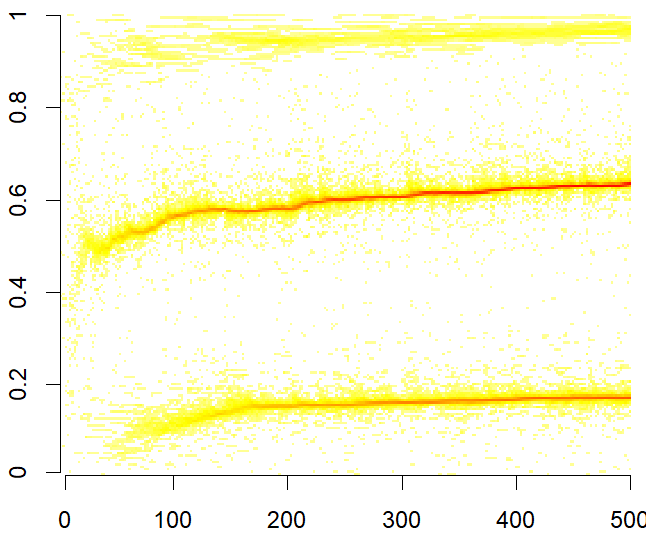}}    &  \makecell{\includegraphics[width = 5 cm]{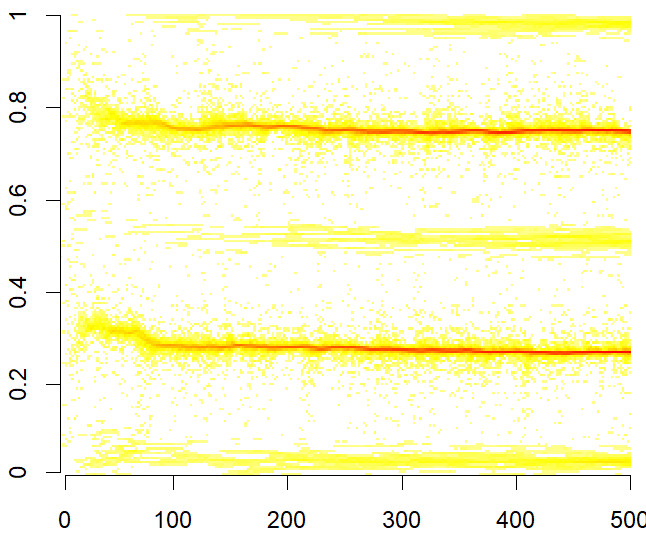}} &  \rotatebox{90}{\hspace{0.0 cm} Agent model}\\

      \rotatebox{90}{\hspace{0 cm}  Opinions} & \makecell{\includegraphics[width = 6 cm]{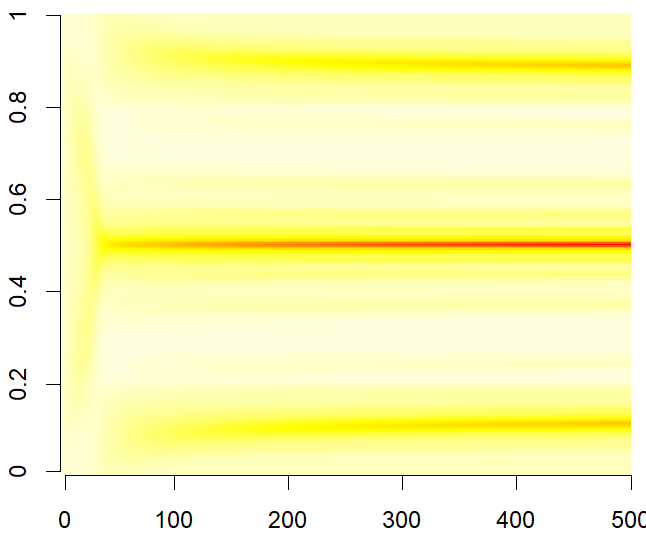}}    &   \makecell{\includegraphics[width = 6 cm]{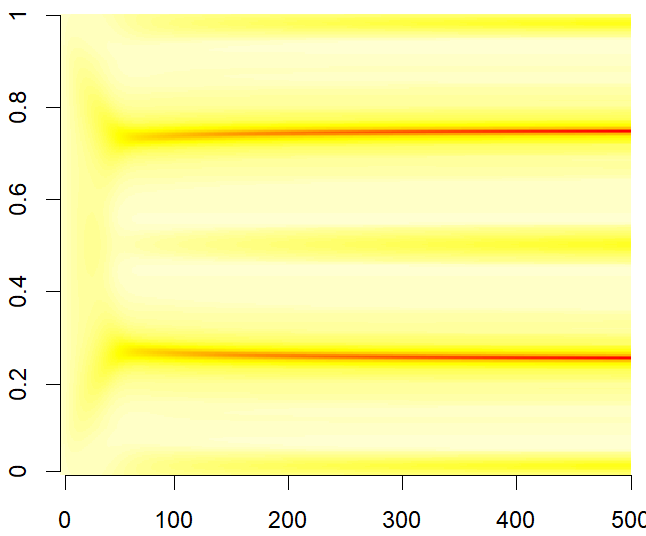}} & \rotatebox{90}{\hspace{0.3 cm} Density model} \\
       &  $t$ / $N_t$ & $t$ / $N_t$ &\\

    \end{tabular}

    \caption{Density of opinions over time for the agent and density models for initial number of agents $N_0 = 5$ and adding $\omega N_0 = 5$ agents at each round (see text for details). }
    \label{fig:examplesN05}
\end{figure}

The upper part of figure \ref{fig:examplesN05} shows examples of runs of the model for an initial number of agents $N_0 = 5$, and adding $\omega N_0 = 5$ agents at each round. On the bottom part, we show the density model for $\omega = 1$. Indeed, in this density model, the initial number $N_0$ cannot be taken into account, as the initial state is a perfect uniform distribution in any case.

\subsection{Configurations of clusters when the confidence bound $\epsilon$ varies}

\begin{figure}[tp]
    \centering
   Cluster positions for the continuous model with standard interactions
    \begin{tabular}{m{0.3cm}cc}
    & &  \\
    &  Growing ($\omega = 1)$  &  Not growing ($\omega = 0$)  \\
     \rotatebox{90}{\hspace{0 cm} Opinions} & \makecell{\includegraphics[width = 6 cm]{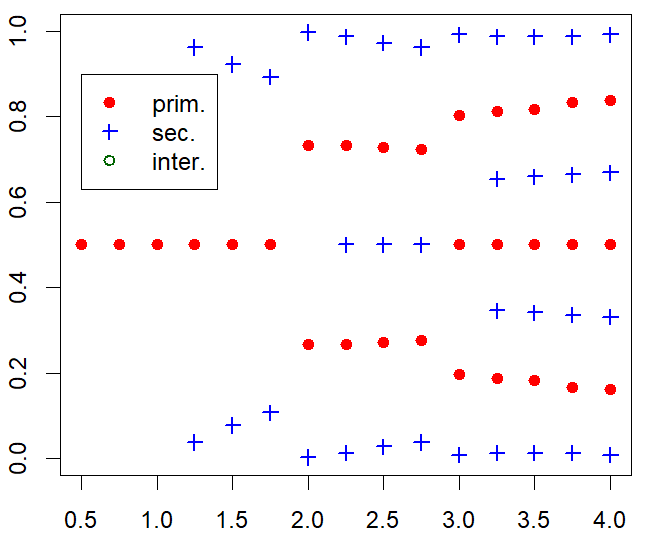}}    &  \makecell{\includegraphics[width = 6 cm]{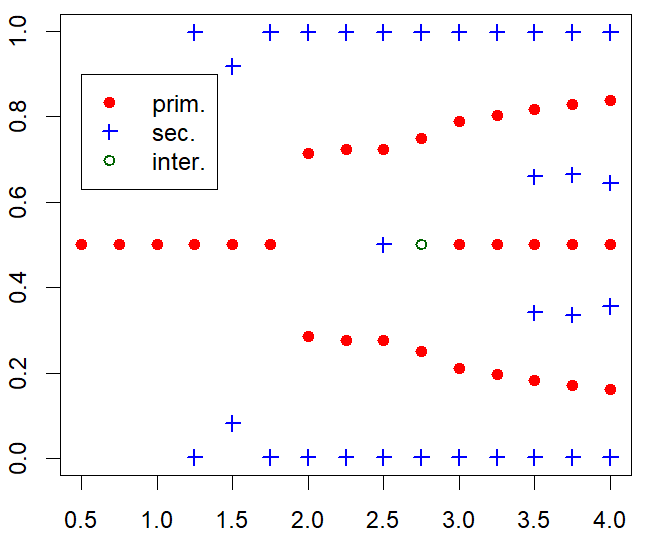}} \\
    & $\frac{1}{2 \epsilon}$ & $\frac{1}{2 \epsilon}$

    \end{tabular}

    \caption{Positions of clusters from the density model for both growing and non-growing cases for different values $\frac{1}{2 \epsilon}$. The primary (prim. in the legend) clusters are such that their effective weight is higher than 0.6, the secondary clusters (sec. in the legend) are such that their effective weight is lower than 0.2. The intermediate clusters (inter. in the legend) are such that their effective weight is between 0.2 and 0.6. See the appendix for the definition of the effective weight of a cluster. }
    \label{fig:clusterPos}
\end{figure}

Figure \ref{fig:clusterPos} shows the positions of the clusters for the continuous approximation model when growing ($\omega = 1$) and not growing ($\omega = 0$), as a function of $\frac{1}{2 \epsilon}$. The figure distinguishes between primary, secondary and intermediate clusters. The distinction is based on the effective weight of the cluster (see the caption of figure \ref{fig:clusterPos}). The effective weight of a cluster is defined in the appendix of this paper.

Overall, the growing and non-growing cases yield similar patterns of cluster positions. A few differences are, however, noticeable: for $\frac{1}{2 \epsilon} = 2.25$, a central secondary cluster is detected only in the growing case. Similarly, for $\frac{1}{2 \epsilon} = 3.25$, the growing case shows two secondary clusters more than the non-growing case. For $\frac{1}{2 \epsilon} = 2.75$, the central cluster is secondary for the growing case, while it is intermediate for the non-growing case. Finally, the positions of the clusters are a bit different. These positions change more sharply at the transitions in the growing case than in the non-growing one.

The configurations of clusters obtained with the density model are references for the agent model. For a given value of the confidence bound $\epsilon$, depending on the other parameters ($N_0$ and $\omega$), the agent model can yield more or less often the same configuration as the density model.

\begin{figure}[h]
    \centering
  Average cluster positions and numbers for the agent model with standard influence
    \begin{tabular}{m{0.3cm}cc}
    & &  \\
    &  $N_0 = 1000$, $\omega = 0.01$, $T = 200$  &   $N_0 = 5$, $\omega = 1$, $T = 500$   \\
     \rotatebox{90}{\hspace{0 cm} Opinions} & \makecell{\includegraphics[width = 6 cm]{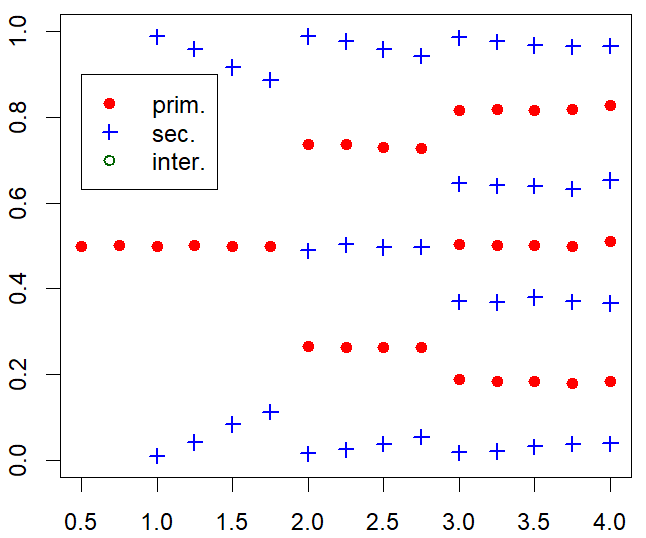}}    &  \makecell{\includegraphics[width = 6 cm]{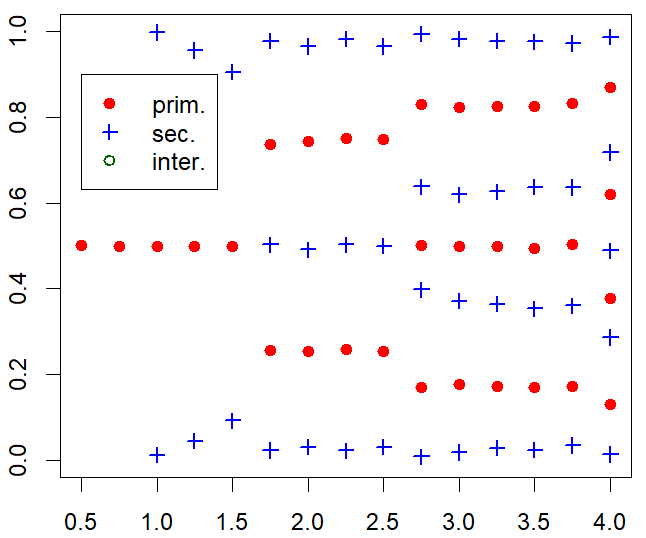}} \\
     & & \\
      & Primary clusters & Secondary clusters \\
          \rotatebox{90}{\hspace{0.0 cm}  cluster nb} & \makecell{\includegraphics[width = 6 cm]{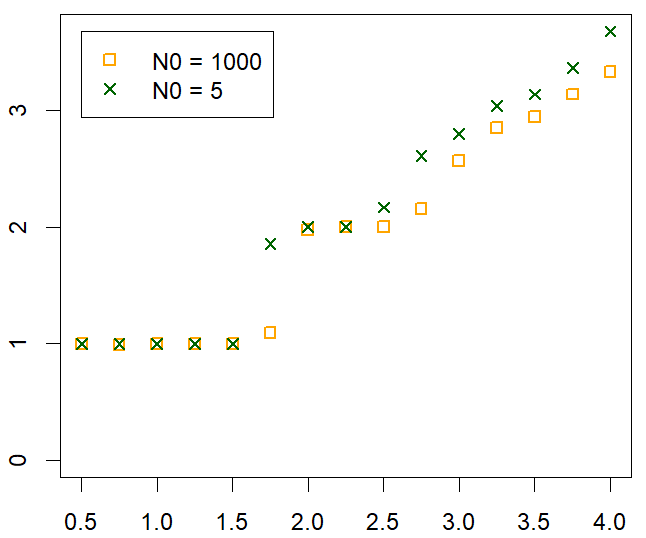}}    &  \makecell{\includegraphics[width = 6 cm]{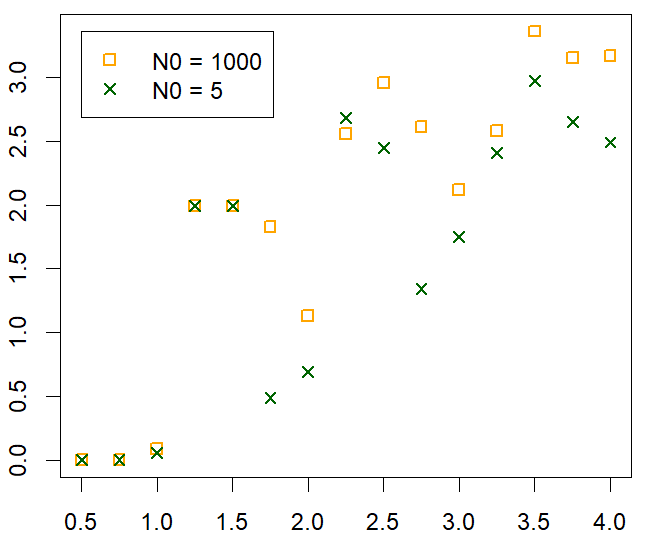}} \\
    & $\frac{1}{2 \epsilon}$ & $\frac{1}{2 \epsilon}$

    \end{tabular}

   \caption{Positions of clusters (top panels) and average number of clusters (bottom panels) from the agent model for an initial number of agents $N_0 = 1000$, $\omega = 0.01$ and after $T = 200$ iterations (left panel) and $N_0 = 5$, $\omega = 1$ and after $T = 500$ iterations (right panel).  }
    \label{fig:clusterPosABM}
\end{figure}

The top panels of figure \ref{fig:clusterPosABM} show the average positions of the clusters for the agent-based model in two cases. In one case (right panel), the initial number $N_0 = 1000$ and the average number of agents added at each iteration is $\omega N_0 = 10$, and the cluster positions are measured after $T = 200$ iterations (on average $3000$ agents). In the other case (left panel), the initial number $N_0 = 5$ and the average number of agents added at each iteration is $\omega N_0 = 5$, and the cluster positions are measured after $T = 500$ iterations (on average $2500$ agents). In each case, the figure shows the average of the most frequent configuration (defined by the number of primary clusters) over 100 replicas for each value of $\epsilon$. The definition of the clusters with their effective weight is the same as in Figure \ref{fig:clusterPos}. Comparing these results with the results from the density model yields the following main observations:
\begin{itemize}
    \item The average positions of clusters for $N_0 = 1000$ and $\omega = 0.01$ (left panel of figure \ref{fig:clusterPosABM}) are very close to the result obtained with the continuous growing model, even though $\omega = 1$ for that continuous case. The only noticeable difference is that the agent model shows additional secondary clusters at the transitions between 2 and 3, and 3 and 4 primary clusters (for $\frac{1}{2\epsilon} \in \{2, 3\}$),
    \item The transitions take place at lower values of $\frac{1}{2\epsilon}$ when the initial number of agents is $N_0 = 5$ (right panel) than when $N_0 = 1000$. Indeed, for $N_0 = 5$, three transitions are observable: between 1 and 2 primary clusters for $\frac{1}{2\epsilon} \in [1.5, 1.75]$, between 2 and 3 primary clusters for $\frac{1}{2\epsilon} \in [2.5, 2.75]$, and between 3 and 4 primary clusters for $\frac{1}{2\epsilon} \in [3.75, 4]$. In contrast, for $N_0 = 1000$, only two transitions are observable: between 1 and 2 primary clusters for $\frac{1}{2\epsilon} \in [1.75, 2]$, and between 2 and 3 primary clusters for $\frac{1}{2\epsilon} \in [2.75, 3]$.
\end{itemize}

These observations suggest that, in general, a low initial number of agents $N_0$ leads to a higher number of primary clusters in the most frequent configurations (for the same average number of agents $\omega N_0$ added at each iteration).

This observation is completed by the bottom panels of figure \ref{fig:clusterPosABM}, showing the average number of clusters for all the configurations. Indeed, the left panel shows that the average number of primary clusters is a bit higher for $N_0 = 5$ than for $N_0 = 1000$. However, on the contrary, the right panel shows that the number of secondary clusters is generally higher for $N_0 = 1000$. This suggests that, for $N_0 = 5$, when the primary clusters are more numerous, they tend to be more often too close to each other to allow the emergence of secondary clusters.

\section{Simulations with the smooth influence function}

\begin{figure}[tp]
    \centering
        Cluster positions for the density model with smooth influence
    \begin{tabular}{m{0.3cm}cc}
    & &  \\
    &  Growing ($\omega = 1)$  &  Not growing ($\omega = 0$)  \\
     \rotatebox{90}{\hspace{0 cm} Opinions} & \makecell{\includegraphics[width = 6 cm]{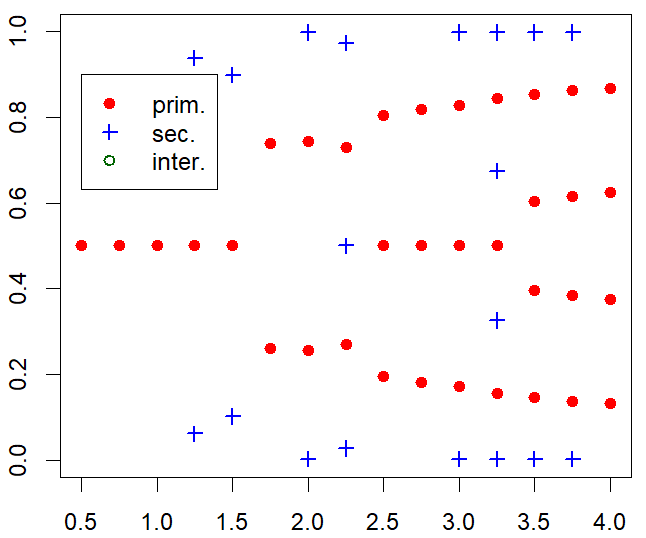}}    &  \makecell{\includegraphics[width = 6 cm]{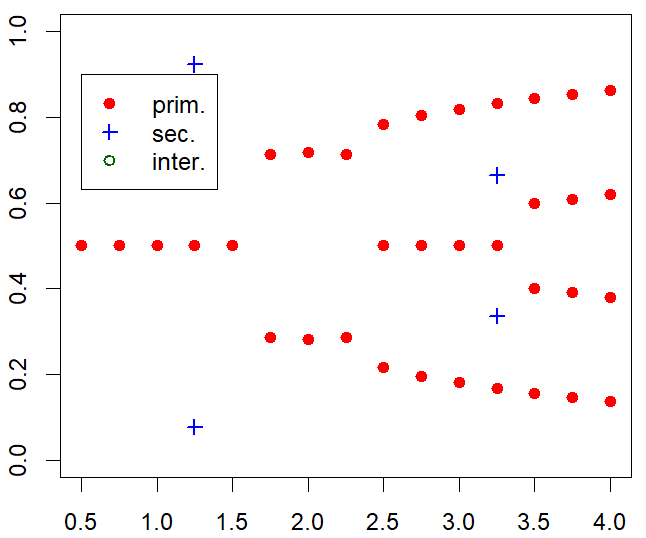}} \\
    & $\frac{1}{2 \epsilon}$ & $\frac{1}{2 \epsilon}$\\
    && \\
    \end{tabular}

    Examples for $\epsilon = 0.2$ ($\frac{1}{2 \epsilon} = 2.5$) \\

    \begin{tabular}{m{0.3cm}cc}
    && \\
    & Growing ($\omega = 1$) & Not growing ($\omega = 0$)  \\
     \rotatebox{90}{\hspace{0 cm} Opinions} & \makecell{\includegraphics[width = 6 cm]{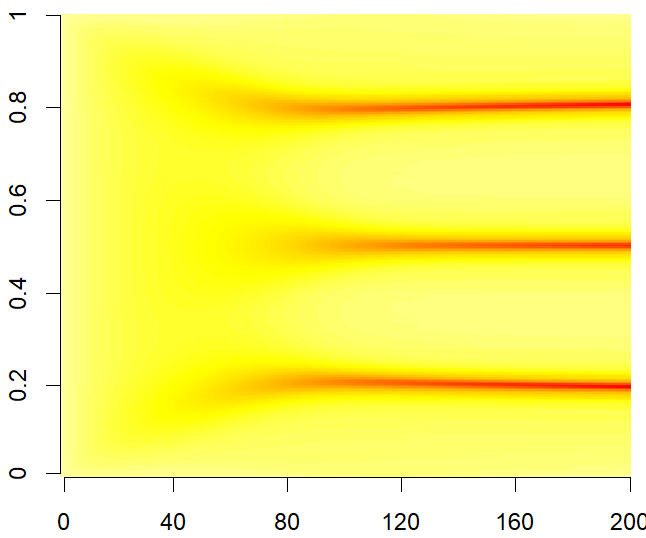}}    &  \makecell{\includegraphics[width = 6 cm]{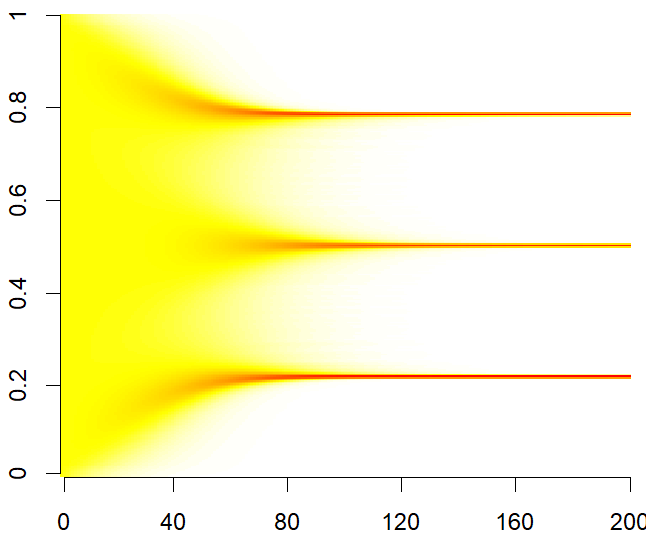}} \\
    & $t/N_t$ & $t/N_t$
    \end{tabular}

    \caption{Positions of clusters for the smooth influence from the density model for both growing and non-growing cases for different values $\frac{1}{2 \epsilon}$. The clusters are defined like previously and computed at the simulation's last iteration ($T = 200$). }
    \label{fig:clusterPosSmmoth}
\end{figure}

The top panels of figure \ref{fig:clusterPosSmmoth} show the positions of the clusters as a function of  $\epsilon$ for the density smooth model, for the case of growing (left) and non-growing (right) population. In the same columns but at the bottom of the same figure, it is shown the simulation for $\epsilon = 0.2$. It is striking that the secondary clusters are almost absent in the non-growing case. There are some additional secondary clusters in the growing case, which are located at the extremes of the opinion interval. Moreover, the transitions between configurations of numbers of primary clusters take place at lower values of $\frac{1}{2\epsilon}$ than in the standard influence.

These results suggest that the secondary clusters found in the growing case with smooth influence are generated by a different process than the one generating the secondary clusters in the non-growing case, which seems mainly due to the discontinuity of the standard influence function. We summary our results in table \ref{table:1}.

\begin{table}[h!]
\begin{center}
\begin{tabular}{ |c|c|c| }
 \hline
  & Fixed population & Growing population \\ \hline \hline
 Standard influence & Discontinuity &  Discontinuity + Arrival of new agents to influence-free areas \\  \hline
 Smooth influence &  & Arrival of new agents to influence-free areas \\
 \hline
\end{tabular}
\end{center}
\caption{Processes generating secondary clusters}
\label{table:1}
\end{table}

\begin{figure}[tp]
    \centering
  Average cluster positions and numbers for the agent model with smooth influence
    \begin{tabular}{m{0.3cm}cc}
    & &  \\
    &  $N_0 = 1000$, $\omega = 0.01$, $T = 200$  &   $N_0 = 5$, $\omega = 1$, $T = 500$   \\
     \rotatebox{90}{\hspace{0.0 cm} Opinions} & \makecell{\includegraphics[width = 6 cm]{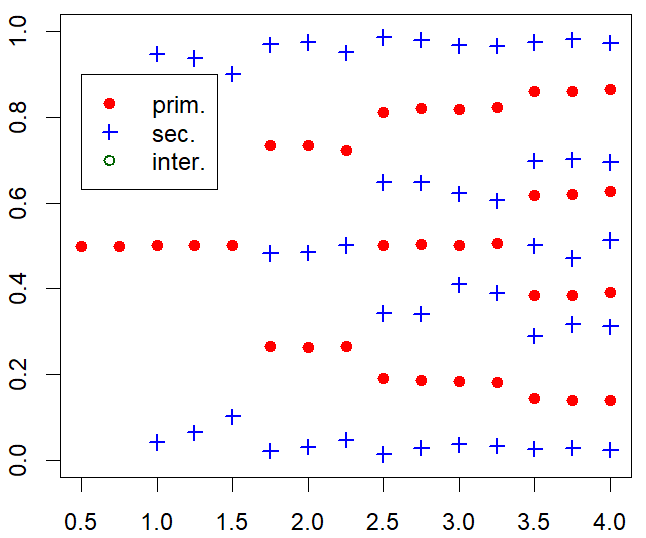}}    &  \makecell{\includegraphics[width = 6 cm]{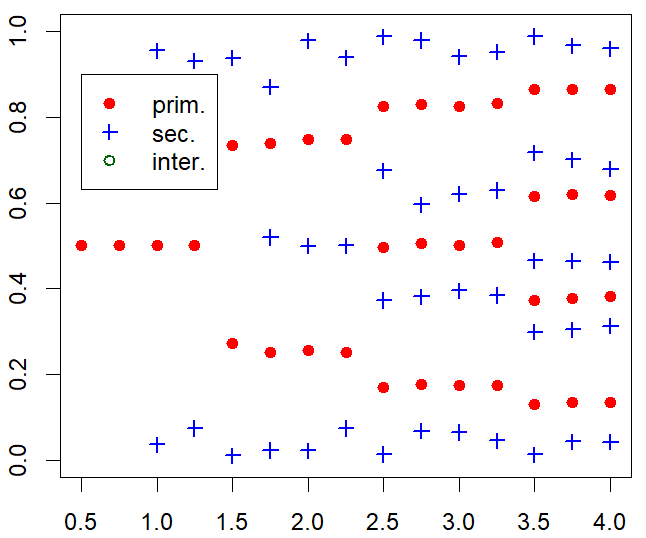}} \\
     & & \\
      & Primary clusters & Secondary clusters \\
          \rotatebox{90}{\hspace{0.0 cm} Cluster nb} & \makecell{\includegraphics[width = 6 cm]{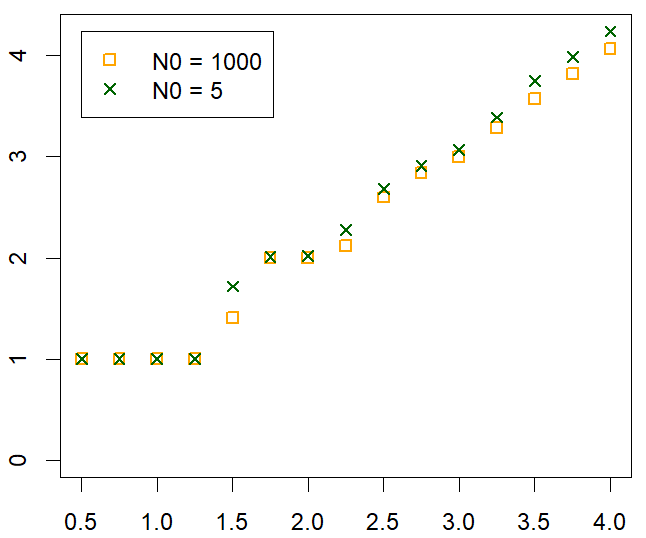}}    &  \makecell{\includegraphics[width = 6 cm]{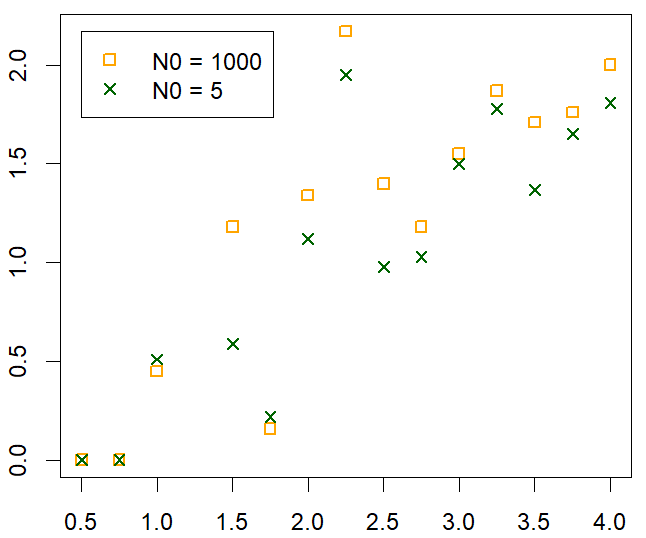}} \\
    & $\frac{1}{2 \epsilon}$ & $\frac{1}{2 \epsilon}$

    \end{tabular}

    \caption{Positions of clusters (top panels) and average number of clusters (bottom panels) from the agent model for an initial number of agents $N_0 = 1000$, $\omega = 0.01$ and after $T = 200$ iterations (left panel) and $N_0 = 5$, $\omega = 1$ and after $T = 500$ iterations (right panel).  }
    \label{fig:clusterPosABMsmooth}
\end{figure}

Figure \ref{fig:clusterPosABMsmooth} is the equivalent of figure \ref{fig:clusterPosABM}, replacing the standard influence with the smooth one. It shows the average positions of the clusters for the most frequent configuration over 100 replicas of simulations of the agent model in the top panels. The average number of primary and secondary clusters is shown in the bottom panels. The left top panel shows the results of the simulations for an initial number of agents $N_0 = 1000$ and adding on average $\omega N_0 = 10$ agents at each iteration. The right top panel shows the results for $N_0 = 5$ and adding on average $\omega N_0 = 5$ agents at each iteration. The positions of the clusters (top panels) are the average for the most frequent configuration defined by the number of primary clusters over 100 replicas (for each value of $\epsilon$). The average number of clusters (bottom line) is performed on all configurations. The left panel shows the average number of primary clusters, and the right panel shows the average number of secondary clusters. The definition of the clusters with their effective weight is the same as in Figure \ref{fig:clusterPos}.

The average positions of the clusters of the most frequent configurations are similar in these top panels. The main difference is that the transition between 1 and 2 primary clusters takes place for $\frac{1}{2\epsilon} \in [1.5, 1.75]$ for $N_0 = 1000$, like the continuous model, and for $\frac{1}{2\epsilon} \in [1.25, 1.5]$ for $N_0 = 5$. The transitions from 2 to 3, and from 3 to 4 primary clusters take place in the same intervals $\frac{1}{2\epsilon} \in [2.25, 2.5]$ and $\frac{1}{2\epsilon} \in [3.25, 3.5]$, respectively, like in the continuous model.

The positions of the secondary clusters are not regular, probably because, in general, only a small part of the secondary clusters that are visible on the figures are present in each configuration participating in the average. For instance, for $N_0 = 1000$ and $\frac{1}{2\epsilon} = 3.75$, the average number of secondary clusters is $0.65$. Therefore, in most simulations, there is at most only one secondary cluster. As a consequence, each position of a secondary cluster is averaged on a small sample, which explains the irregularities.

The average numbers of primary clusters in all the configurations (left bottom panel) are very similar for $N_0 = 1000$ and $N_0 = 5$, the number being slightly higher for $N_0 = 5$. Note that the number grows almost linearly for $\frac{1}{2\epsilon} > 2.5$.

There are more differences in the average numbers of secondary clusters (right bottom panel). Most of the time, the average number of secondary clusters is smaller for $N_0 = 5$, which seems difficult to explain only by the small difference of the number of primary clusters. Hence this point may require further investigations. Overall, the number of secondary clusters is significantly lower with the smooth influence than with the standard one, confirming our claim that the secondary clusters, in the smooth case, are emerging only because of the regular arrival of new agents in locations of the opinion interval that are not under the influence of a primary cluster.

\section{Discussion}
This section summarises our results and discusses the remaining open questions.

The behaviour of the bounded confidence model on a growing population is significantly different from the noisy version of the model because the distribution of agents tends to a density with several fixed peaks (primary and secondary), while in the noisy bounded confidence, this is generally not the case \cite{Pineda2009,Pineda2011}. From our simulations, the model can reach different steady states in density, like the model on a fixed population, depending on random events taking place in the initial phase of the simulation. However, checking the model on very long simulations is difficult as the number of agents keeps growing.

The most striking novelty of the model on a growing population is the emergence of secondary clusters, which are larger than when the population is fixed, and more importantly, they appear systematically, which is not the case in the fixed population. Nevertheless, when considering the density models, the maps of the primary and secondary clusters when the confidence bound varies are similar in the growing and non-growing population cases. This may suggest that the secondary clusters appear with a similar process whether the population is growing or not, the growing population simply reinforcing an existing process.

The results obtained with smooth interactions suggest otherwise. Indeed, in this case, the model on a fixed population shows almost no secondary cluster, while the secondary clusters are still present in the simulations of the model on a growing population, albeit less numerous.

This suggests that when the population is fixed, the secondary clusters are generated by the discontinuity of the standard influence and when the population is growing, they come from another process (summarised in table \ref{table:1}). Our work suggests the following explanations:
\begin{itemize}
    \item The standard influence function shows its maximum effect at the border of the attraction basin of a developing cluster. As a result of this strong attraction, the density of opinion is likely to be depleted inside the basin in the vicinity of its border. Therefore, the opinions located just outside the attraction basin can become too far from the opinions within the basin, and are thus not attracted. This situation does not occur with the smooth influence function, because the opinions located at the border of the attraction basin move slowly and have more chances to remain close enough to attract opinions beyond the border while the opinions are progressively gathering into a peak.
    \item With both standard or smooth influence, the primary clusters are often further than $2 \epsilon$ from each other or further than $\epsilon$ from the limit of the opinion interval. This leaves some regions of the opinion interval which are not under the influence of a primary cluster. When the population is growing, the opinions that are regularly recruited in these areas, progressively feed secondary clusters.
    \item These secondary clusters can maintain themselves only if they are significantly smaller than their neighbouring primary clusters. Indeed, the regular arrival of new opinions in a shared zone of influence is likely to generate a lot of interactions back and forth between the clusters that bring them closer and closer until they ultimately merge. However, suppose that one of the clusters is much smaller than the other. In that case, an opinion arriving in their common zone of influence is very likely to be attracted only by the bigger cluster. Therefore, the back-and-forth interactions are very unlikely, and both clusters can maintain themselves. Moreover, their difference in size keeps increasing because the bigger cluster is more likely to attract the new agents (this process resembles the "preferential attachment" in social networks).
    \item Finally, we showed that varying the initial number of agents in the population leads to significantly different results. In particular, when this number is small (e.g. 5 to 10), the model shows a first phase during which the position of the primary clusters can change significantly, possibly because of interactions with secondary clusters. These phenomena would be interesting subjects for future investigations.
\end{itemize}

\section{Appendix: effective weight of clusters}

The method for computing the effective weight of the clusters in a distribution of opinions involves three main steps:
\begin{itemize}
    \item Computing a smooth distribution with Gaussian kernel operator of variance about $\alpha \epsilon$ (we generally choose $\alpha = 0.1$),
    \item Selecting the local maxima of the smooth distribution that dominates the distribution in a vicinity of $\beta \epsilon$ (we generally choose $\beta = 0.5$),
    \item Computing the effective weight of each local maximum as its weight multiplied by the effective number of clusters.
\end{itemize}

We now describe each step in more details.

\subsection{Computing a smooth distribution with a Gaussian Kernel}

Let $A = (a_1,..., a_N)$ be the distribution of opinions. Let $(x_1,...,x_p)$ be such that for $i \in (1,...,p)$, $x_i$ is the middle of interval $[\frac{i-1}{p},\frac{i}{p}] $:
\begin{align}
    x_i = \frac{i - 0.5}{p}.
\end{align}
For any number $x$, let the Gaussian function $G(x_i, x)$ be (with $\alpha = 0.1$ a parameter):
\begin{align}
    G(x_i, x) = \exp\left( -\left(\frac{x - x_i}{\alpha \epsilon}\right)^2\right).
\end{align}

For each $x_i$, the smoothed distribution value is defined as:
\begin{align}
    S(x_i) = \sum_{j = 1}^N G(x_i, a_j).
\end{align}

This smoothing erases strong irregularities of the histogram of opinions and provides a more accurate view of the respective weights of the clusters than when associating the clusters to the maxima of the opinion histogram.

\begin{figure}[tp]
    \centering
    \begin{tabular}{cc}
    Distribution (red) smoothed (blue) & Effective weights for 100 runs \\
     \includegraphics[width = 7 cm]{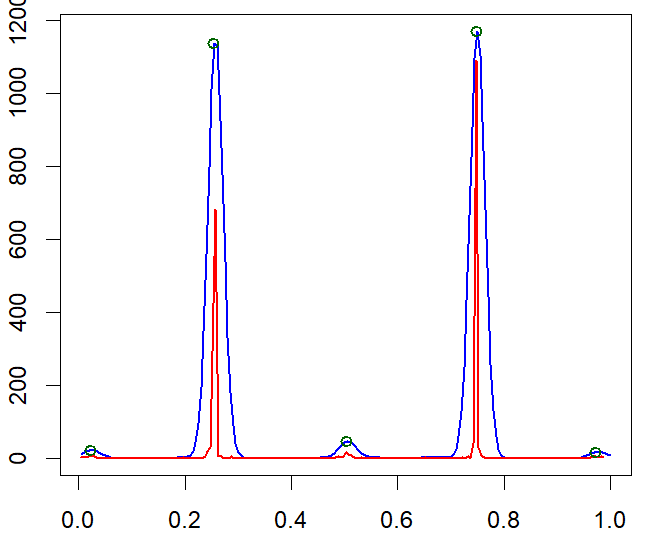}    &  \includegraphics[width = 7 cm]{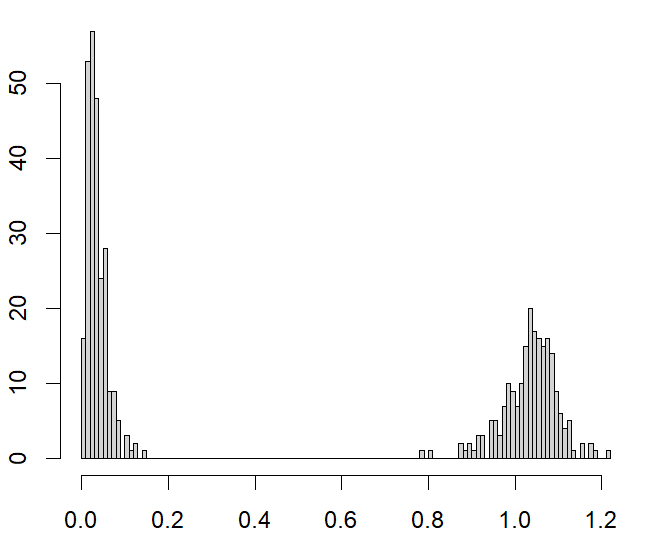}\\
    Opinions & Opinions
    \end{tabular}

    \caption{On the left panel, an illustration of the method for identifying the clusters. The red curve is the distribution of the opinions, the blue curve is the smoothed distribution, the dark green points are the detected local maxima. The effective weights of these clusters are in the order from left to right: $0.02$, $1.02$, $0.04$, $1.05$, $0.01$. The right panel shows the distribution of the values of the cluster effective weights for 100 simulations for $\epsilon = 0.2$, $N_0 = 5$, $\omega = 1$ at $T = 100$ rounds. }
    \label{fig:clusters}
\end{figure}

\subsection{Selecting local maximums of the smooth distribution and computing the effective weight of each of them}

The couple $(x_i, S(x_i))$ defines a local maximum of the smoothed distribution if, for all $j \neq i$ such that $|x_i - x_i| < \beta \epsilon$, $S(x_j) < S(x_i)$.

\subsection{Computing the effective weight associated with a local maximum}

Let $(x_{i_1},...,x_{i_m})$ be the set of values defining the local maximums. Then $w_{i_j}$, the weight of the maximum is:

\begin{align}
    w_{i_j} = \frac{S(x_{i_j})}{\sum_{k=1}^m S(x_{i_k})}.
\end{align}

From weights, we compute $n_g$ the effective number of clusters \cite{Laakso1979} as follows:

\begin{align}
    n_e = \frac{1}{\sum_{k=1}^m w^2_{i_k} }.
\end{align}

If there are $n$ clusters of the same weight $\frac{1}{n}$, then $n_e = n$, and if there are major clusters and minor clusters, $n_e$ tends to be close to the number of major clusters.

The effective $W_{i_j}$ of maximum $i_j$ is finally defined as:

\begin{align}
    W_{i_j} = n_e w_{i_j}.
\end{align}

In general, the primary clusters define local maxima with an effective weight around 1, while for the secondary clusters the corresponding effective weight is lower than 0.2. In the example shown on the right pane of figure \ref{fig:clusters}, there are no clusters of effective weight comprised between 0.2 and 0.7. Therefore, the approach discriminates well between secondary (low effective weight) and primary (high effective weight) clusters.
\bibliography{references.bib}
\end{document}